\documentclass{ptephy_v1}

\preprintnumber{XXXX-XXXX} 

\usepackage{amsmath,amssymb}
\usepackage{bm}
\usepackage{braket}
\usepackage[subrefformat=parens]{subcaption}
\usepackage{graphics} 

\begin{document}

\title{Analytical and numerical assessment of accuracy
  of the approximated nuclear symmetry energy in the Hartree-Fock theory}





\author{Y. Tsukioka and H. Nakada*
}
\affil{Department of Physics, Graduate School of Science, Chiba University, Yayoi-cho 1-33, Inage, Chiba 263-8533, Japan \email{nakada@faculty.chiba-u.jp}}


\begin{abstract}
The nuclear symmetry energy is defined
by the second derivative of the energy per nucleon
with respect to the proton-neutron asymmetry,
and is sometimes approximated by the energy difference
between the neutron matter and the symmetric matter.
Accuracy of this approximation is assessed analytically and numerically,
within the Hartree-Fock theory using effective interactions.
By decomposing the nuclear-matter energy,
the relative error of each term is expressed analytically;
it is constant or is a single-variable function determined by the function type.
The full errors are evaluated for several effective interactions,
by inserting values of the parameters.
Although the errors stay within $10\%$ up to twice of the normal density
irrespective of the interactions,
at higher densities accuracy of the approximation
significantly depends on the interactions.
\end{abstract}

\subjectindex{xxxx, xxx}

\maketitle

\section{Introduction}   

The nuclear symmetry energy,
which is defined as the second derivative of the energy per nucleon
with respect to the proton-neutron asymmetry, is an important quantity
for prediction of masses of neutron-rich nuclei
and structure of neutron stars.
Calculation of the symmetry energy, however, is not always easy.
Analytical approaches are impossible
unless energy of the system is given by a twice-differentiable function
with respect to the asymmetry,
while a numerical evaluation requires precise calculation
of energy of the system.
For this reason, the symmetry energy is sometimes approximated
as energy difference between the pure neutron matter and symmetric matter.
Conversely, this approximation of the symmetry energy corresponds
to the quadratic approximation of the neutron-matter energy
with respect to the asymmetry,
in its estimation from the symmetric-matter energy.

As Bethe originally pointed out~\cite{NM},
this quadratic approximation of energy is applicable to small asymmetry
but its validity is not obvious for systems with large asymmetry
like the neutron matter.
Accuracy of this approximation has been studied in several works.
Chen \textit{et al.} investigated this approximation at the saturation density
with a modified Gogny interaction~\cite{Ho}.
This approximation was examined in the Brueckner-Hartree-Fock theory
with the Paris potential by Bombaci and Lombardo~\cite{ANM},
and with the AV18 plus three-nucleon force by Zuo \textit{et al.}~\cite{3BF}.
Drischler \textit{et al.} also studied it
in the chiral effective field theory \cite{EFT}.
It should be mentioned that Wellenhofer \textit{et. al.} pointed out
that the approximation is not valid at certain temperatures~\cite{HFB}.

We reinvestigate accuracy of the approximation of the symmetry energy
at zero temperature within the Hartree-Fock theory,
by using effective interactions
which have been tested for structure of finite nuclei.
This enables analytical arguments,
by which origin of the errors can be examined term by term.
Then the errors can be numerically estimated
by inserting values of the parameters.
In practice we adopt six effective interactions;
the Skyrme interactions SkM$^\ast$~\cite{SkM*} and SLy4~\cite{SLy4},
the Gogny interactions D1S~\cite{D1S} and D1M~\cite{D1M},
and the M3Y-type interactions M3Y-P6 and M3Y-P7~\cite{M3Y-P}.

This paper is organized as follows.
In Sect.~2, we decompose the nuclear-matter energy, the symmetry energy,
and its error.
From Sect.~3 to 5, we evaluate the relative errors of individual terms.
Section~3 is devoted to the errors
whose analytical forms are independent of the effective Hamiltonian.
The errors specific to the Skyrme
and finite-range (Gogny and M3Y-type) effective interactions
are analytically investigated in Sect.~4 and 5, respectively.
The numerical results of errors for symmetry energy are displayed
and discussed in Sect.~6.
Section~7 provides a summary of the paper.

\section {Decomposition of energies and errors}

In this section, we shall give expressions
of the symmetry energy $a_{t}(\rho)$
and its approximation ${\tilde a}_{t}(\rho)$,
by decomposing them into several terms.
Correspondingly, the errors of the symmetry energy can be decomposed as well.

\subsection{Effective Hamiltonian}

In this paper,
we handle the homogeneous nuclear matter with the spin degeneracy
in the Hartree-Fock theory.

Because of the homogeneity, the single-particle wave function
is written as a plane wave, 
\begin{equation}\label{spwf}
 \varphi _{\bm{k}\sigma\tau}(\bm{r})
 = \frac{1}{\sqrt{\Omega}} e^{i\bm{k}\cdot\bm{r}} \chi_{\sigma} \chi_{\tau}\,,
\end{equation}
where $\bm{k}$ denotes the momentum and $\Omega$ is the volume of the system,
for which we shall take $\Omega \rightarrow \infty$ later.
$\chi_{\sigma}$ ($\chi_{\tau}$) is the spin (isospin) wave function.

The effective Hamiltonian of the system
consists of the kinetic energy and the effective interaction,
\begin{equation}\label{Ham}
 H = K + V\,; \quad
 K=\sum_{i} \frac{\bm{p}_{i}^{2}}{2M}\,,\quad V = \sum_{i<j} v_{ij}\,,
\end{equation}
where $i$ and $j$ are indices of nucleons. 
The effective interaction $V$ is comprised of the central term,
the LS term and the tensor (TN) term.
In the homogeneous matter,
the LS term and the TN term can be neglected.
We therefore treat only the central term,
which may contain usual two-body interaction and density-dependent interaction.
For the latter we assume contact form.
We then have
\begin{equation}\label{int}
\begin{split}
 v_{12} &= v_{12}^{(\mathrm{C})} + v_{12}^{(\mathrm{DD})}\,;\\
&\quad v _{12}^{(\mathrm{C})} = \sum _n \bigl(t_n^{(\mathrm{SE})}P_\mathrm{SE}
 +t_n^{(\mathrm{TE})}P_\mathrm{TE}+t_n^{(\mathrm{SO})}P_\mathrm{SO}
 +t_n^{(\mathrm{TO})}P_\mathrm{TO}\bigr)f_n(r_{12})\,,\\
&\quad v _{12}^{(\mathrm{DD})}
= t_\rho^{(\mathrm{SE})} P_\mathrm{SE}\,
 [\rho(\bm{r}_1)]^\alpha\,\delta(\bm{r}_{12})
+ t_\rho^{(\mathrm{TE})} P_\mathrm{TE}\,
 [\rho(\bm{r}_1)]^\beta\,\delta(\bm{r}_{12})\,,
\end{split}
\end{equation}
where $f_n(r_{12})$ is an appropriate function of $r_{12}=|\bm{r}_{12}|$
with $\bm{r}_{12}=\bm{r}_1-\bm{r}_2$.
In this paper, we treat the Skyrme, the Gogny, and the M3Y-type interactions.
Correspondingly, $f_n(r_{12})$ is the $\delta$-function
(with and without momentum-dependence), the Gauss function,
or the Yukawa function.
The effective interactions have range parameters,
and the index $n$ distinguishes the range.
All kinds of $t$ in the above equation are strength parameters.
$P_\mathrm{TE}$, $P_\mathrm{TO}$, $P_\mathrm{SE}$, and $P_\mathrm{SO}$
are projection operators to Singlet-Even, Triplet-Even, Singlet-Odd,
and Triplet-Odd two-particle states,
and they are related to the spin exchange operator $P_{\sigma}$
and the isospin exchange operator $P_{\tau}$ as follows,
\begin{equation}  
\begin{aligned}
P_\mathrm{SE}&=\frac{1-P_\sigma}{2}\frac{1+P_\tau}{2}\,,\quad
P_\mathrm{TE}=\frac{1+P_\sigma}{2}\frac{1-P_\tau}{2}\,,\\
P_\mathrm{SO}&=\frac{1-P_\sigma}{2}\frac{1-P_\tau}{2}\,,\quad
P_\mathrm{TO}=\frac{1+P_\sigma}{2}\frac{1+P_\tau}{2}\,.\end{aligned}
\end{equation}
We can rewrite $v_{12}^{(\mathrm{C})}$ as follows,
\begin{equation}\label{vc}
\begin{split}
v_{12}^{(\mathrm{C})}&= \sum_n
 (t_n^{(\mathrm{W})}+t_n^{(\mathrm{B})}P_{\sigma}-t_n^{(\mathrm{H})}P_{\tau}
 -t_n^{(\mathrm{M})}P_{\sigma}P_{\tau})f_n(r_{12})\,;\\
&\qquad t_n^{(\mathrm{W})}=\bigl( t_n^{(\mathrm{SE})}+t_n^{(\mathrm{TE})}
 +t_n^{(\mathrm{SO})}+t_n^{(\mathrm{TO})}\bigr)/4\,,\\ 
&\qquad t_n^{(\mathrm{B})}=\bigl(-t_n^{(\mathrm{SE})}+t_n^{(\mathrm{TE})}
 -t_n^{(\mathrm{SO})}+t_n^{(\mathrm{TO})}\bigr)/4\,,\\
&\qquad t_n^{(\mathrm{H})}=\bigl(-t_n^{(\mathrm{SE})}+t_n^{(\mathrm{TE})}
 +t_n^{(\mathrm{SO})}-t_n^{(\mathrm{TO})}\bigr)/4\,,\\ 
&\qquad t_n^{(\mathrm{M})}=\bigl( t_n^{(\mathrm{SE})}+t_n^{(\mathrm{TE})}
 -t_n^{(\mathrm{SO})}-t_n^{(\mathrm{TO})}\bigr)/4\,.
\end{split}
\end{equation}

\subsection{Decomposition of energy per nucleon}

With the wave function of Eq.~\eqref{spwf}
and the effective Hamiltonian of Eqs.~\eqref{Ham}, \eqref{int},
the energy per nucleon ${\cal E}$ can be expressed
in the following manner,
as was derived in Ref.~\cite{M3Y}.
We consider a function ${\cal W}$ as
\begin{equation}
{\cal W}(k_{\tau},k_{\tau'})= \int_{k_1\le k_{\tau}}d^3k_1
\int_{k_2\le k_{\tau'}}d^3k_2\,\tilde{f}_n(|\bm{k}_{12}-\bm{k}'_{12}|),
\end{equation}
where $\tilde{f}(q)$ signifies the Fourier transform of $f(r)$;
$\tilde{f}(q)=\int d^3r\,f(r)e^{-i\bm{q}\cdot\bm{r}}$. 
$\bm{k}_{12}$ and $\bm{k}'_{12}$ represent relative momenta
of the initial and the final states.
Since $\bm{k}_{12}=\bm{k}'_{12}$ in the Hartree (\textit{i.e.}, direct) term
and $\bm{k}_{12}=-\bm{k}'_{12}$ in the Fock (\textit{i.e.}, exchange) term,
contributions of these terms to the energy
is represented by the function ${\cal W}$ as follows,
\begin{equation}\label{w-func}
\begin{split}
{\cal W}^{\rm{H}}_n(k_{\tau},k_{\tau'})&= \int_{k_1\le k_{\tau}}d^3k_1
\int_{k_2\le k_{\tau'}}d^3k_2\,\tilde{f}_n(0)
=\frac{16\pi^2}{9}k_{\tau}^3k_{\tau'}^3\tilde{f}_n(0)\quad
\mbox{(Hartree term)}\,,\\
{\cal W}^{\rm{F}}_n(k_{\tau},k_{\tau'})&= \int_{k_1\le k_{\tau}}d^3k_1
\int_{k_2\le k_{\tau'}}d^3k_2\,\tilde{f}_n(2k_{12}) \\
&=8\pi^2\biggl[\int_{0}^{|k_{\tau'}-k_{\tau}|/2}dk_{12}\,
\frac{16}{3}k_{\tau}^3 k_{12}^2\tilde{f}(2k_{12})
+\int_{|k_{\tau'}-k_{\tau}|/2}^{(k_{\tau'}+k_{\tau})/2} dk_{12}
\Bigl\{-\frac{1}{2}(k_{\tau'}^2-k_{\tau}^2)^2 k_{12}\\
&\qquad +\frac {8}{3}(k_{\tau}^3+k_{\tau'}^3)k_{12}^2
-4(k_{\tau}^2+k_{\tau'}^2)k_{12}^{3}
+\frac {8}{3}k_{12}^5 \Bigr\} \tilde{f}_n(2k_{12}) \biggr]
\quad\mbox{(Fock term)}\,.
\end{split}
\end{equation}

We can express the total energy by using these functions.
Under the spin degeneracy, the total energy is represented by
\begin{equation}\label{E}
\begin{aligned}
E&=\frac{\Omega}{10\pi^2M }(k_\mathrm{p}^5+k_\mathrm{n}^5)\\
&\quad +\frac{\rho^\alpha\Omega}{36\pi^4}\,
 \bigl[t_\rho^{(\mathrm{SE})}\rho^\alpha(k_\mathrm{p}^6+k_\mathrm{n}^6)
 +(t_\rho^{(\mathrm{SE})}\rho^\alpha+3t_\rho^{(\mathrm{TE})}\rho^\beta)
 k_\mathrm{p}^3 k_\mathrm{n}^3\,\bigr]\\
&\quad +\frac{\Omega}{(2\pi)^6}\sum _{n}
 \bigl[(2t_n^{(\mathrm{W})}+t_n^{(\mathrm{B})}-2t_n^{(\mathrm{H})}
 -t_n^{(\mathrm{M})})
 \bigl\{{\cal W}^{\rm{H}}_n(k_\mathrm{p},k_\mathrm{p})
 +{\cal W}^{\rm{H}}_n(k_\mathrm{n},k_\mathrm{n})\bigr\}\\
&\qquad\qquad\quad\quad +2(2t_n^{(\mathrm{W})}+t_n^{(\mathrm{B})})\,
 {\cal W}^{\rm{H}}_n(k_\mathrm{p},k_\mathrm{n})\\
&\qquad\qquad\quad\quad+(2t_n^{(\mathrm{M})}+t_n^{(\mathrm{H})}
 -2t_n^{(\mathrm{B})}-t_n^{(\mathrm{W})})
\bigl\{{\cal W}^{\rm{F}}_n(k_\mathrm{p},k_\mathrm{p})
 +{\cal W}^{\rm{F}}_n(k_\mathrm{n},k_\mathrm{n})\bigr\}\\
&\qquad\qquad\quad\quad +2(2t_n^{(\mathrm{M})}+t_n^{(\mathrm{H})})\,
 {\cal W}^{\rm{F}}_n(k_\mathrm{p},k_\mathrm{n})\bigr]\,,
\end{aligned}
\end{equation}
where $k_\mathrm{p}$ ($k_\mathrm{n}$) denotes
the Fermi momentum of protons (neutrons),
and they are connected with the density $\rho_\mathrm{p}$ ($\rho_\mathrm{n}$)
or with the total density $\rho$ and the asymmetry $\eta_t$ as
\begin{equation}
\begin{split}
k_\mathrm{p}=(3\pi^2\rho_\mathrm{p})^{1/3}
=\Bigl\{\frac{3\pi^2}{2} \rho (1-\eta_t)\Bigr\}^{1/3}\,,\quad
& k_\mathrm{n}=(3\pi^2\rho_\mathrm{n})^{1/3}
=\Bigl\{\frac{3\pi^2}{2} \rho (1+\eta_t)\Bigr\}^{1/3}\,;\\
&\qquad \rho=\rho_\mathrm{p}+\rho_\mathrm{n}\,,\quad
\eta_t=\frac{\rho_\mathrm{n}-\rho_\mathrm{p}}{\rho}\,.
\end{split}
\end{equation}
The energy per nucleon ${\cal E}=E/A$ is acquired
by dividing $E$ by the nucleon number $A=\rho \Omega$.
We here decompose ${\cal E}$ into the kinetic term (${\cal E}_\mathrm{K}$),
the density-dependent term (${\cal E}_\mathrm{DD}$),
the Hartree term between like nucleons (${\cal E}_\mathrm{HO}$)
and between unlike nucleons (${\cal E}_\mathrm{HX}$),
the Fock term between like nucleons (${\cal E}_\mathrm{FO}$)
and between unlike nucleons (${\cal E}_\mathrm{FX}$),
\begin{equation}\label{e}
\begin{aligned}
{\cal E}(\rho,\eta_t)&={\cal E}_{\mathrm{K}} + {\cal E}_{\mathrm{DD}}
+ \sum_n\bigl({\cal E}_{\mathrm{HO}n} + {\cal E}_{\mathrm{HX}n}
+ {\cal E}_{\mathrm{FO}n} + {\cal E}_{\mathrm{FX}n}\bigr)\,;\\
 {\cal E}_\mathrm{K}
 &=\frac{1}{10\pi^2 M\rho}(k_\mathrm{p}^5+k_\mathrm{n}^5)\,,\\
 {\cal E}_\mathrm{DD} &= \frac{1}{36\pi^4}
 \bigl[t^{(\mathrm{SE})}_\rho \rho ^{\alpha-1}
 (k_\mathrm{p}^6+k_\mathrm{n}^6+k_\mathrm{p}^3 k_\mathrm{n}^3)
 + 3t^{(\mathrm{TE})}_\rho\rho^{\beta -1}k_\mathrm{p}^3 k_\mathrm{n}^3\bigr]\,,
 \\
{\cal E}_{\mathrm{HO}n}&=\frac{1}{(2\pi)^6\rho}
 \cdot\bigl(2t_n^{(\mathrm{W})}+t_n^{(\mathrm{B})}-2t_n^{(\mathrm{H})}
 -t_n^{(\mathrm{M})}\bigr)
 \bigl\{{\cal W}^{\rm{H}}_n(k_\mathrm{p},k_\mathrm{p})
 +{\cal W}^{\rm{H}}_n(k_\mathrm{n},k_\mathrm{n})\bigr\}\,,\\
{\cal E}_{\mathrm{HX}n}&=\frac{1}{(2\pi)^6\rho }
 \cdot 2(2t_n^{(\mathrm{W})}+t_n^{(\mathrm{B})})\,
 {\cal W}^{\rm{H}}_n(k_\mathrm{p},k_\mathrm{n})\,,\\
{\cal E}_{\mathrm{FO}n}&=\frac{1}{(2\pi)^6\rho}
 \cdot\bigl(2t_n^{(\mathrm{M})}+t_n^{(\mathrm{H})}-2t_n^{(\mathrm{B})}
 -t_n^{(\mathrm{W})}\bigr)
 \bigl\{{\cal W}^{\rm{F}}_n(k_\mathrm{p},k_\mathrm{p})
 +{\cal W}^{\rm{F}}_n(k_\mathrm{n},k_\mathrm{n})\bigr\}\,,\\
{\cal E}_{\mathrm{FX}n}&=\frac{1}{(2\pi)^6\rho}
 \cdot 2(2t_n^{(\mathrm{M})}+t_n^{(\mathrm{H})})\,
 {\cal W}^{\rm{F}}_n(k_\mathrm{p},k_\mathrm{n})\,.
\end{aligned}
\end{equation} 

\subsection{Decomposition of symmetry energy}

The symmetry energy $a_t(\rho)$ is defined
by the second-order derivative of ${\cal E}$
with respect to the asymmetry $\eta_t$,
\begin{equation}
a_t(\rho) := \left.\frac{1}{2}\frac{\partial^2{\cal E}}{\partial\eta_t^2}
 \right|_{\eta_t=0}\,.
\end{equation}
Let us denote the $\nu$-th order partial derivative
of a function ${\cal F}$ with respect to $\eta_t$
by ${\cal F}^{(\nu)} = (\partial^\nu/\partial\eta_t^\nu){\cal F}$.
Corresponding to the decomposition of ${\cal E}$,
we also decompose the symmetry energy as follows,
\begin{equation}\label{ai}
\begin{aligned}
a_t(\rho) &= \left.\frac{1}{2}{\cal E}^{(2)}\right|_{\eta_t=0}
=a_\mathrm{K} + a_\mathrm{DD} + \sum _{n}\big(a_{\mathrm{HO}n} + a_{\mathrm{HX}n}
+ a_{\mathrm{FO}n} + a_{\mathrm{FX}n}\big)\,;\\
&\quad
a_i=\left.\frac{1}{2}{\cal E}_i^{(2)}\right|_{\eta_t= 0}\quad
\mbox{($i=\mathrm{K}, \mathrm{DD}, \mathrm{HO}n, \mathrm{HX}n,
\mathrm{FO}n, \mathrm{FX}n$)}.
\end{aligned}
\end{equation}
Formulas for calculating each $a_i$ are given in Appendix~\ref{symeng}.
Note that $\left.{\cal W}^{\rm{H}}_n(k_\mathrm{p},k_\mathrm{p})^{(\nu)}
\right|_{k_\mathrm{p}=k_\mathrm{F}}
\neq \left.{\cal W}^{\rm{H}}_n(k_\mathrm{p},k_\mathrm{n})^{(\nu)}
\right|_{k_\mathrm{p}=k_\mathrm{n}=k_\mathrm{F}}$.

\subsection{Approximation of symmetry energy and its error}

The symmetry energy $a_t(\rho)$ is often approximated
by the difference of ${\cal E}$ between $\eta_t=1$ and $\eta_t=0$, 
\begin{equation}\label{at}
\begin{aligned}
\tilde{a}_t(\rho) &:= {\cal E}(\rho,\eta_t=1)-{\cal E}(\rho,\eta_t=0)\\
 &= a_t(\rho) + \sum_{\nu=2}^{\infty} 
 \left.\frac{{\cal E}(\rho,\eta_t)^{(2\nu)}}{(2\nu)!}\right|_{\eta_t=0}\,.
\end{aligned}
\end{equation}
This coincides with the quadratic approximation of ${\cal E}$
with respect to $\eta_t$,
\begin{equation}
{\cal E}(\rho,\eta_t)\approx {\cal E}(\rho,\eta_t=0) + a_t(\rho)\,\eta_t^{2}\,.
\end{equation}
Notice that ${\cal E}$ is an even function of $\eta_t$
under the charge symmetry.

We are now ready to consider accuracy of the approximation of Eq.~\eqref{at}.
As measures of the approximation,
we will estimate the absolute error $\delta_t={\tilde a}_t-a_t$
and the relative error $\Delta_t = ({\tilde a}_t-a_t)/a_t$.
In correspondence to each term of Eq.~\eqref{ai}
($a_i(\rho)$; $i=\mathrm{K}, \mathrm{DD}, \mathrm{HO}n, \mathrm{HX}n,
\mathrm{FO}n, \mathrm{FX}n$), we define its approximated value
${\tilde a}_i(\rho)$ by
\begin{equation}
{\tilde a}_i(\rho) := {\cal E}_i(\rho,\eta_t=1)-{\cal E}_i(\rho,\eta_t=0)\,.
\end{equation}
Then the relative error of each term $\Delta_i=({\tilde a}_i-a_i)/a_i$
can be estimated analytically, as shown below.
The full errors $\delta_t$ and $\Delta_t$ are expressed by using $\Delta_i$,
\begin{equation}
\begin{aligned}
\delta_t &={\tilde a}_t-a_t=\sum_i \Delta_i\,a_i\,,\\
\Delta_t &= \frac{{\tilde a}_t-a_t}{a_t}=\sum_i \Delta_i\frac{a_i}{a_t}\,.
\end{aligned}
\end{equation}
Thus, through $\Delta _{i}$,
the error of the symmetry energy can be examined term by term.
We shall estimate $\Delta_i$ in the following three sections.

\section{Terms with constant relative errors}

The relative errors ($\Delta_i$) for the kinetic term,
the density-dependent term, and the Hartree term of central force
are constant, or even vanish, independent of the density. 
In this section we discuss $\Delta_i$ for these terms.

\subsection{Kinetic term}

We first discuss the kinetic energy term
in the exact and the approximated symmetry energy,
$a_\mathrm{K}$ and ${\tilde a}_\mathrm{K}$. 
The $2\nu$-th order coefficient of kinetic energy per nucleon $(1 \le \nu)$
can be derived inductively,
\begin{equation}
  \left.\frac{{\cal E}_\mathrm{K}^{(2\nu)}(\rho,\eta_t)}{(2\nu)!}\right|_{\eta_t=0}
  =\frac{k_\mathrm{F}^5}{5\pi^2M\rho}\cdot\frac{1}{(2\nu)!}
    \prod_{i=0}^{2\nu-1}\frac{5-3i}{3}\,;\quad
  k_\mathrm{F}=\Bigl(\frac{3\pi^2\rho}{2}\Bigr)^{1/3}\,.
\end{equation}
The relative error of kinetic term in the symmetry energy
turns out to be constant, independent of $\rho$,
\begin{equation}\label{DK}
\Delta_\mathrm{K}=\frac{{\tilde a}_\mathrm{K}-a_\mathrm{K}}{a_\mathrm{K}}
= \sum_{\nu=2}^{\infty}\left.\frac{{\cal E}_\mathrm{K}^{(2\nu)}/(2\nu)!}
  {{\cal E}_\mathrm{K}^{(2)}/2}\right|_{\eta_t=0}
= \sum_{\nu=2}^{\infty}\left[\prod_{i=1}^{\nu-1}\frac{(6i-5)(6i-2)}{(6i+3)(6i+6)}\right]
= 0.057\,.
\end{equation}
In Eq.~(6) of Ref.\cite{HFB},
the result equivalent to Eq.~\eqref{DK} is presented,
$({\tilde a}_\mathrm{K}-a_\mathrm{K})/{\tilde a}_\mathrm{K}
=\Delta_\mathrm{K}/(1+\Delta_\mathrm{K})=0.054$.

\subsection{Density-dependent term}

For the density-dependent contact term, we have
\begin{equation}
 {\tilde a}_\mathrm{DD}=a_\mathrm{DD}\,,\quad \Delta_\mathrm{DD}=0\,.
\end{equation}
This is because ${\cal E}_\mathrm{DD}$ can be written up to the quadratic term
with respect to $\eta_t$,
\begin{equation}
{\cal E}_\mathrm{DD}(\rho,\eta_t)=\frac{1}{8}
\bigl[\,t_\rho^{(\mathrm{SE})}\rho^{\alpha+1}(3+\eta_t^2)
  +3t_\rho^{(\mathrm{TE})}\rho^{\beta+1}(1-\eta_t^2)\,\bigr]\,.
\end{equation}

\subsection{Hartree-term}

The Hartree term of the central force in the energy per nucleon
also depends quadratically on $\eta_t$.
In practice, Eq.~\eqref{w-func} yields
\begin{equation} 
\begin{aligned}
  {\cal W}^\mathrm{H}_n(k_\mathrm{p},k_\mathrm{p})
  &=4\pi^6\rho^2(1-\eta_t)^2\,\tilde{f}_n(0)\,,\\
  {\cal W}^\mathrm{H}_n(k_\mathrm{n},k_\mathrm{n})
  &=4\pi^6\rho^2(1+\eta_t)^2\,\tilde{f}_{n}(0)\,,\\
  {\cal W}^\mathrm{H}_n(k_\mathrm{p},k_\mathrm{n})
  &=4\pi^6\rho^2(1-\eta_t^2)\,\tilde{f}_{n}(0)\,,
\end{aligned}
\end{equation}
leading to $\Delta_{\mathrm{HO}n}=\Delta_{\mathrm{HX}n}=0$.

\section{Skyrme interaction}

We next discuss the terms depending on the function types of the interaction,
through which full expression of errors of the symmetry energy is obtained.

The Skyrme interaction has momentum-dependent terms,
additional to Eq.~\eqref{int},
while some of the terms in Eq.~\eqref{vc} become equivalent
to exchange terms of others.
The interaction is expressed, instead of Eq.~\eqref{int},
\begin{equation}
\begin{split}
v_{12} &= t_0(1 + x_0 P_\sigma)\,\delta(\bm{r}_{12})\\ 
&\quad +\frac{1}{2}t_1(1+x_1 P_\sigma)
 \left[\bm{p}_{12}^{\prime 2}\,\delta(\bm{r}_{12})
  +\delta(\bm{r}_{12})\,\bm{p}_{12}^{2} \right]
+t_2(1+x_2 P_\sigma)\bm{p}_{12}^{\prime}\cdot\delta(\bm{r}_{12})\,\bm{p}_{12}\\
&\quad +t_\rho^{(\mathrm{SE})}P_\mathrm{SE},[\rho(\bm{r}_1)]^\alpha\,\delta(\bm{r} _{12})
+ t_\rho^{(\mathrm{TE})}P_\mathrm{TE}\,[\rho(\bm{r}_1)]^\beta\,\delta(\bm{r} _{12})\,,
\end{split}
\end{equation}
Here $\bm{p}_{12}=(\nabla_1-\nabla_2)/(2i)$,
and $\bm{p}_{12}^{\prime}$ is the hermitian conjugate of $\bm{p}_{12}$
acting on the left.
For the $t_0$ term that has $f(r_{12})=\delta(\bm{r}_{12})$,
we have
\begin{equation}
  {\cal W}^{\rm H}_0(k_\tau,k_{\tau'})={\cal W}^{\rm F}_0(k_\tau,k_{\tau'})=\frac{16\pi^2}{9}k_\tau^3 k_{\tau'}^3\,.
\end{equation}
For the $t_1$ and $t_2$ terms, 
$\frac{1}{2}(\bm{p}_{12}^{\prime 2}\,\delta(\bm{r}_{12})
+\delta(\bm{r}_{12})\,\bm{p}_{12}^2)$
and $\bm{p}_{12}^\prime\cdot\delta(\bm{r}_{12})\,\bm{p}_{12}$ yield~\cite{M3Y}
\begin{equation}
\begin{aligned}
  {\cal W}^{\rm H}_1(k_\tau,k_{\tau'})={\cal W}^{\rm F}_1(k_\tau,k_{\tau'})
  &= \frac{4\pi^2}{15} k_\tau^3 k_{\tau'}^3(k_\tau^2+k_{\tau'}^2)\quad\mbox{($t_1$ term)}\,,\\
  {\cal W}^{\rm H}_2(k_\tau,k_{\tau'})=-{\cal W}^{\rm F}_2(k_\tau,k_{\tau'})
  &= \frac{4\pi^2}{15} k_\tau^3 k_{\tau'}^3(k_\tau^2+k_{\tau'}^2)\quad\mbox{($t_2$ term)}\,,
\end{aligned}
\end{equation}
respectively.

Let us define 
\begin{equation}
\begin{split}
{\cal W}^{\rm o}_\mathrm{c}=\frac{16\pi^2}{9}(k_\mathrm{p}^6 + k_\mathrm{n}^6)\,,&\quad
{\cal W}^{\rm x}_\mathrm{c}=\frac{16\pi^2}{9}k_\mathrm{p}^3 k_\mathrm{n}^3\,,\\
{\cal W}^{\rm o}_\mathrm{p}=\frac{8\pi^2}{15}(k_\mathrm{p}^8 + k_\mathrm{n}^8)\,,&\quad
{\cal W}^{\rm x}_\mathrm{p}=\frac{8\pi^2}{15}k_\mathrm{p}^3 k_\mathrm{n}^3
 (k_\mathrm{p}^2 + k_\mathrm{n}^2)\,,
\end{split}
\end{equation}
where the superscript ${\rm o}$ (${\rm x})$ indicates
like- (unlike-) nucleon contribution
and the subscript $\mathrm{c}$ $(\mathrm{p})$ corresponds
to the $t_0$ ($t_{1}$ or $t_{2}$) term.
Then the energy per nucleon ${\cal E}$ is decomposed as follows.
\begin{equation}\label{ep}
\begin{split}
  {\cal E} &= {\cal E}_\mathrm{K}+{\cal E}_\mathrm{c}+{\cal E}_\mathrm{pO}
  +{\cal E}_\mathrm{pX}+{\cal E}_\mathrm{DD}\,;\\
  &\qquad {\cal E}_\mathrm{c}= \frac{1}{(2\pi)^6 \rho}
  \bigl(t^{\rm{o}}_\mathrm{c}\,{\cal W}^{\rm{o}}_\mathrm{c}
    +t^{\rm{x}}_\mathrm{c}\,{\cal W}^{\rm{x}}_\mathrm{c} \bigr)\,,\\
  &\qquad {\cal E}_\mathrm{pO}= \frac{1}{(2\pi)^6 \rho}\
    t^{\rm{o}}_\mathrm{p}\,{\cal W}^{\rm{o}}_\mathrm{p}\,,\quad
  {\cal E}_\mathrm{pX}= \frac{1}{(2\pi)^6 \rho}
    t^{\rm{x}}_\mathrm{p}\,{\cal W}^{\rm{x}}_\mathrm{p}\,, 
\end{split}
\end{equation}
where
\begin{equation}
\begin{split}
 t^{\rm o}_\mathrm{c} = t_0(1-x_0)\,,&\quad t^{\rm x}_\mathrm{c} = t_0(2+x_0)\,,\\
 t^{\rm o}_\mathrm{p} = t_1(1-x_1)+3t_2(1+x_2)\,,&\quad
 t^{\rm x}_\mathrm{p} = t_1(2+x_1)+t_2(2+x_2)\,.
\end{split}
\end{equation}
Corresponding to the above decomposition,
the symmetry energy and its error are decomposed into $a_i$ and $\Delta_i$
($i=\mathrm{K},\mathrm{c},\mathrm{pO},\mathrm{pX},\mathrm{DD}$).
$\delta_t$ and $\Delta_t$ are expressed by
\begin{equation}
\begin{split}
  \delta_t &= \tilde{a}_t-a_t = \Delta_\mathrm{K} a_\mathrm{K}
  + \Delta_\mathrm{pO} a_\mathrm{pO} + \Delta_\mathrm{pX} a_\mathrm{pX}\,,\\
  \Delta_t &= \frac{\tilde{a}_t-a_t}{a_t}
  = \Delta_\mathrm{K}\frac{a_\mathrm{K}}{a_t}
  + \Delta_\mathrm{pO}\frac{a_\mathrm{pO}}{a_t}
  + \Delta_\mathrm{pX}\frac{a_\mathrm{pX}}{a_t}\,.
\end{split}
\end{equation}

We have shown, in the previous section,
$\Delta_\mathrm{K}=0.057$ and $\Delta_\mathrm{DD}=0$.
Moreover, we obviously have $\Delta_\mathrm{c}=0$.
For the momentum-dependent terms,
we consider the ratios $\tilde{a}_\mathrm{pO}/a_\mathrm{pO}=1+\Delta_\mathrm{pO}$
and $\tilde{a}_\mathrm{pX}/a_\mathrm{pX}=1+\Delta_\mathrm{pX}$,
where $a_\mathrm{pO}=\left.\frac{1}{2}{\cal E}_\mathrm{pO}^{(2)}\right|_{\eta_t=0}$
and $a_\mathrm{pX}=\left.\frac{1}{2}{\cal E}_\mathrm{pX}^{(2)}\right|_{\eta_t=0}$.
Since
\begin{equation}
\label{wp_2n}
\begin{aligned}
  \left.{\cal W}^{{\rm o}(2\nu)}_\mathrm{p}(\rho,\eta_t)\right|_{\eta_t=0}
  &= \frac{16\pi^2}{15}k_\mathrm{F}^8 \prod_{i=1}^{2\nu}\frac{11-3i}{3}\,,\\
  \left.{\cal W}^{{\rm x}(2\nu)}_\mathrm{p}(\rho,\eta_t)\right|_{\eta_t=0}
  &= \frac{16\pi^2}{15}k_\mathrm{F}^8(3\nu-2)\prod_{i=1}^{2\nu}\frac{11-3i}{3}\,,
\end{aligned}
\end{equation}
the ratios are
\begin{equation}
\begin{split}
\frac{{\tilde a}_\mathrm{pO}}{a_\mathrm{pO}}
&=\left.\frac{\sum_{\nu=1}^{\infty}{\cal E}_\mathrm{pO}^{(2\nu)}/(2\nu)!}
     {{\cal E}_\mathrm{pO}^{(2)}/2}\right|_{\eta_t=0}
     = \frac{9}{20} \sum_{\nu=1}^{\infty}
     \left[\frac{1}{3^{2\nu}(2\nu)!}\prod_{i=1}^{2\nu}(11-3i)\right]
     =0.979\,,\\
\frac{{\tilde a}_\mathrm{pX}}{a_\mathrm{pX}}
&=\left.\frac{\sum_{\nu=1}^{\infty}{\cal E}_\mathrm{pX}^{(2\nu)}/(2\nu)!}
     {{\cal E}_\mathrm{pX}^{(2)}/2}\right|_{\eta_t=0}
     = \frac{9}{20} \sum_{\nu=1}^{\infty}
     \left[\frac{3\nu-2}{3^{2\nu}(2\nu)!}\prod_{i=1}^{2\nu}(11-3i)\right]
     =0.900\,.
\end{split}
\end{equation}
Therefore $\delta_t$ and $\Delta_t$ for the Skyrme interaction is calculated
by the following formulas,
\begin{equation}\label{DtS}
\begin{split}
  \delta_t &= \tilde{a}_t-a_t
  = 0.057a_\mathrm{K}-0.021a_\mathrm{pO}-0.100a_\mathrm{pX}\,,\\
  \Delta_t &= \frac {\tilde{a}_t-a_t}{a_t}
  = 0.057\frac{a_\mathrm{K}}{a_t}-0.021\frac{a_\mathrm{pO}}{a_t}
  -0.100\frac{a_\mathrm{pX}}{a_t}\,.
\end{split}
\end{equation}

\section{Finite-range interactions}
 
In this section, errors of each term in the symmetry energy
are analytically investigated
for finite-range effective interactions.
The Gauss function $f_n(r_{12})=e^{-(\mu_n r_{12})^{2}}$
and the Yukawa function $f_n(r_{12})=e^{-\mu_n r_{12}}/\mu_n r_{12}$
are handled in practice,
which are used in the central channels of the Gogny and M3Y-type interactions.

As we showed in Sect.~3,
the density-dependent term and the Hartree term give no errors.
Then the errors of the approximated symmetry energy can be expressed as follows,
\begin{equation}
\begin{aligned}
  \delta_t={\tilde a}_t-a_t
  &= \Delta_\mathrm{K} a_\mathrm{K} + \sum_n \Delta_{\mathrm{FO}n} a_{\mathrm{FO}n}
  + \sum_n \Delta_{\mathrm{FX}n} a_{\mathrm{FX}n}\,,\\
  \Delta_t=\frac{{\tilde a}_t-a_t}{a_t}
  &= \Delta_\mathrm{K}\frac{a_\mathrm{K}}{a_t}
  + \sum_n\Delta_{\mathrm{FO}n}\frac{a_{\mathrm{FO}n}}{a_t}
  + \sum_n\Delta_{\mathrm{FX}n}\frac{a_{\mathrm{FX}n}}{a_t}\,.
\end{aligned}
\end{equation}
It is recalled that $\Delta_\mathrm{K}$ is constant.
Unlike the Skyrme interaction,
the relative errors of Fock terms in the symmetry energy depend on the density.
However, for a given function form,
each of $\Delta_{\mathrm{FO}n}$ and $\Delta_{\mathrm{FX}n}$
depends only on $k_\mathrm{F}/\mu_n$,
where $1/\mu_n$ is the range parameter.
Therefore, we calculate the relative errors
$\Delta_{\mathrm{FO}n}$ and $\Delta_{\mathrm{FX}n}$
as functions of $k_\mathrm{F}/\mu_n$.
We depict the results in Fig.~1 for the Gauss and the Yukawa functions,
whose analytic expressions are given in Appendix~\ref{Delta_F}.
It is emphasized that these results are determined only by the function type
of the effective interaction, independent of parameters.
As $k_\mathrm{F}/\mu_n$ increases,
$|\Delta_{\mathrm{FO}n}|$ and $|\Delta_{\mathrm{FX}n}|$
tend to deviate from zero.
The longest range of the nucleon-nucleon interaction
should be given by the Compton wavelength of the pion ($1.414\,\mathrm{fm}$).
Corresponding to the range $0\le 1/\mu_n\le 1.414\,\mathrm{fm}$
and the density $0 \leq \rho \leq 0.64\,{\rm fm}^{-3}$
(\textit{i.e.}, $0 \le k_\mathrm{F} \le 2.11\,{\rm fm}^{-1}$),
whose upper bound is about four times of the normal density,
$\Delta_{\mathrm{FO}n}$ and $\Delta_{\mathrm{FX}n}$ are displayed
for $0 \le k_\mathrm{F}/\mu _n\le 3$.
\begin{figure}
\begin{minipage}[h]{0.5\hsize}
  \centering
   \includegraphics[scale=0.6]{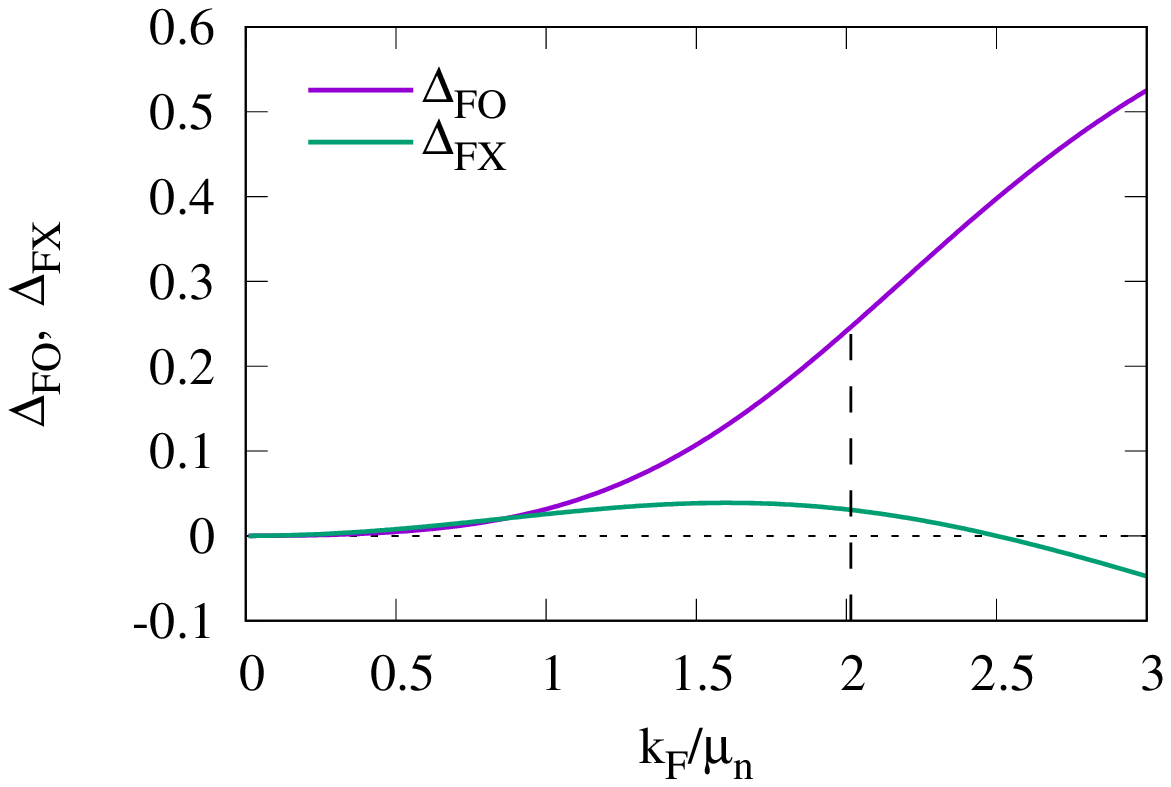}
 \vspace{10pt}
\subcaption{Gauss function} 
\end{minipage}
 \begin{minipage}[h]{0.5\hsize}
  \centering
  \includegraphics[scale=0.6]{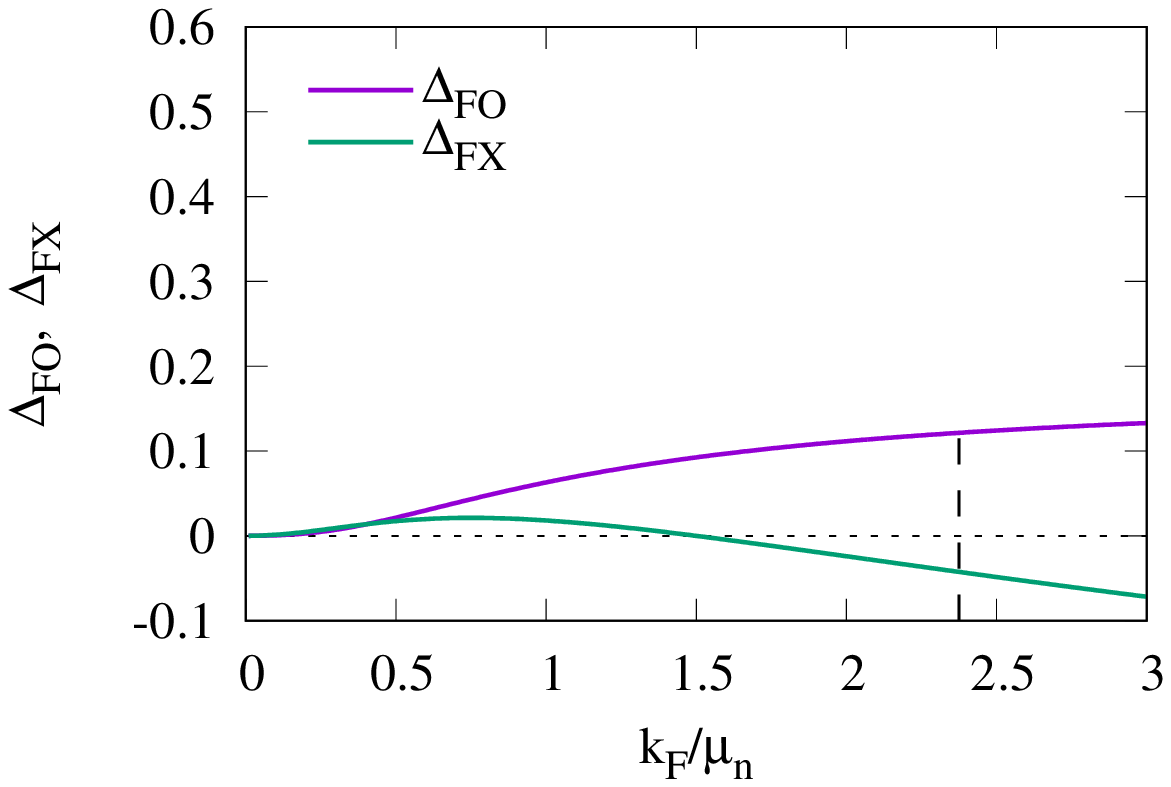}
 \vspace{10pt}
\subcaption{Yukawa function} 
\end{minipage}
 \caption{$k_\mathrm{F}/\mu_n$ dependence of $\Delta_{\mathrm{FO}n}$ and $\Delta_{\mathrm{FX}n}$,
   for (a) the Gauss interaction and (b) the Yukawa interaction.
   The dashed lines are explained in the text.}
\end{figure}

The dashed vertical line in Fig.~1(a) means $k_\mathrm{F}/\mu_n$
for $\rho = 0.32\,{\rm fm}^{-3}$ and $1/\mu_n=1.2\,{\rm fm}$,
which is the longer range in D1S.
We find that $|\Delta_{\mathrm{FX}n}|$ may reach about $0.25$
at $\rho=0.32\,{\rm fm}^{-3}$,
while $|\Delta_{\mathrm{FO}n}|$ is kept within $0.05$ in the full range in Fig.~1(a).

As in Fig.~1(a), the dashed line in Fig.~1(b) means
$k_\mathrm{F}/\mu_n$ for $\rho=0.32{\rm fm}^{-3}$
and the longest range of the M3Y-type interaction $1.414 {\rm fm}^{-1}$.
In contrast to the Gauss function,
$|\Delta_{\mathrm{FO}n}|$ and $|\Delta_{\mathrm{FX}n}|$ stay within $0.1$
in $0 \le \rho \le 0.32 {\rm fm^{-3}}$ for the Yukawa interaction.
Even in the higher density region up to $k_\mathrm{F}/\mu_n=3$,
they are $0.13$ at largest.

\section{Numerical results for several interactions}

In Fig.~2, we show the numerical results of $\Delta_t$ and $\delta_t$
as a function of $\rho$,
which are calculated by inputting the values of the parameters
given in Appendix~\ref{param}. 
\begin{figure}
 \begin{minipage}[t]{0.5\hsize}
  \centering
   \includegraphics[scale=0.6]{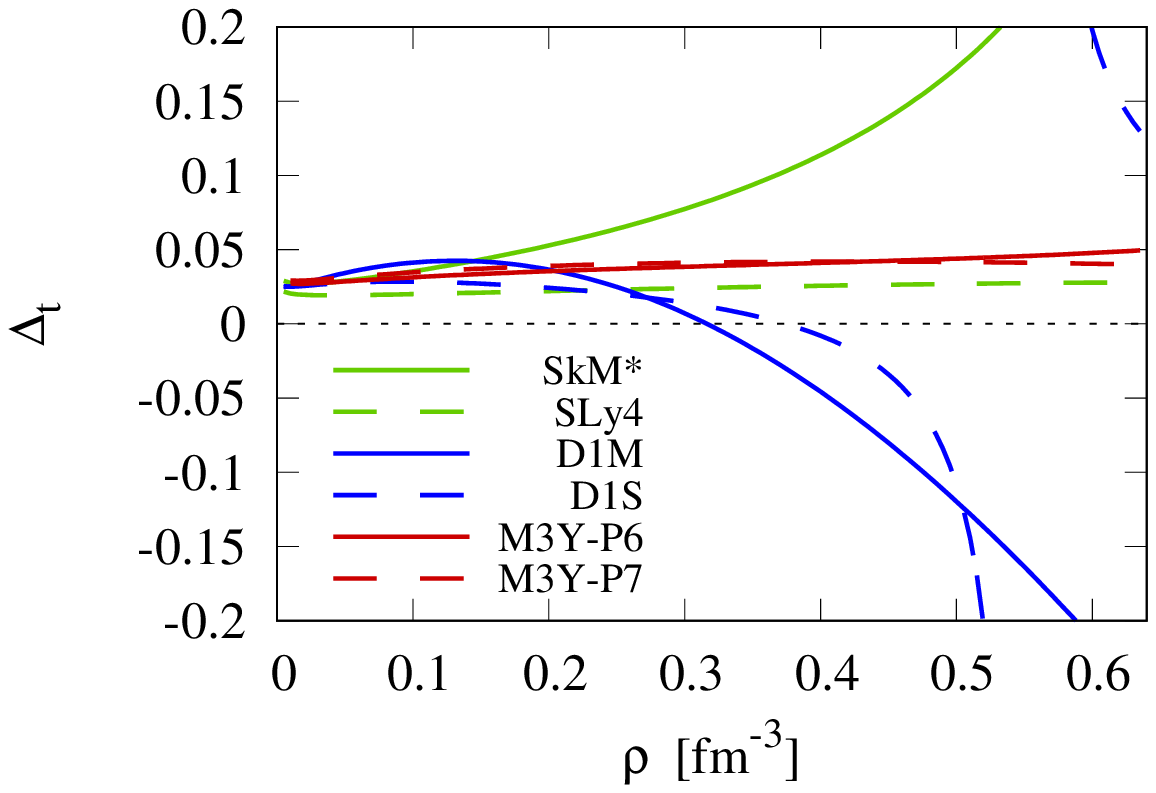}
 \vspace{10pt}
\subcaption{Relative error $\Delta_{t}$}
\end{minipage}
 \begin{minipage}[t]{0.5\hsize}
  \centering
  \includegraphics[scale=0.6]{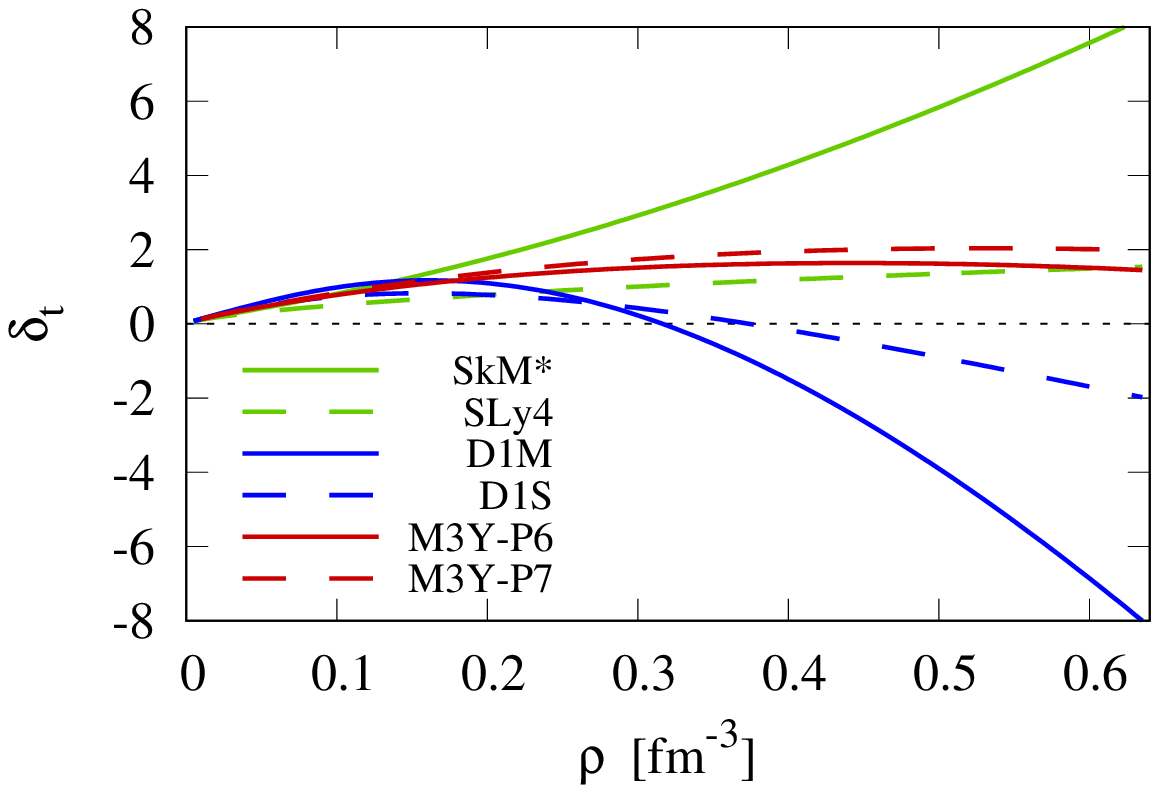}
 \vspace{10pt}
\subcaption{Absolute error $\delta_{t}$} 
\end{minipage}
\caption{Errors for the symmetry energy (a) $\Delta_t$ and (b) $\delta_t$ for several effective interactions.}
\end{figure}
Figure~2(a) reveals interaction-dependence of $\Delta_t$.
In $0<\rho\le 0.32\,{\rm fm}^{-3}$,
we find that the relative error is $|\Delta_t| < 0.1$
for all of the effective interactions.
To be more precise, it is within $|\Delta_t| < 0.05$ except SkM$^\ast$.
In the higher density region of $0.32 \le \rho < 0.64\,{\rm fm}^{-3}$,
$|\Delta_t| < 0.05$ holds for SLy4 and the M3Y-type interactions.
In contrast, $|\Delta_t|$ exceeds $0.20$ for SkM$^\ast$
and the Gogny interactions.
This tendency is also seen for $\delta_t$ in Fig.~2(b).
$|\delta_t|$ of M3Y-P6, M3Y-P7 and SLy4 are significantly smaller
than that of D1M, D1S and SkM$^\ast$.

Interestingly, although SkM$^\ast$ and SLy4 have the same function type,
$\delta_t$ behaves quite differently between them.
Origin of this difference may be accounted for
on the basis of the decomposition in Sect.~4, especially by Eq.~\eqref{DtS}. 
In Fig.~3, $a_\mathrm{K}(\rho)$, $a_\mathrm{pO}(\rho)$ and $a_\mathrm{pX}(\rho)$
for SkM$^\ast$ and SLy4 are presented.
As recognized from Eq.~\eqref{DtS} and Fig.~3,
$\Delta_\mathrm{pX}a_\mathrm{pX}=-0.1a_\mathrm{pX}$ contributes to $\delta_t$
positively.
The contribution ($-a_\mathrm{pX}$) is larger for SkM$^\ast$ than for SLy4.
Moreover, $\Delta_\mathrm{pO}a_\mathrm{pO}=-0.021a_\mathrm{pO}$
contributes negatively
and it tends to cancel the positive contribution of $a_\mathrm{pX}$ for SLy4,
while $a_\mathrm{pO}$ almost vanishes for SkM$^\ast$.
\begin{figure}
\begin{minipage}[t]{0.5\hsize}
  \centering
   \includegraphics[scale=0.6]{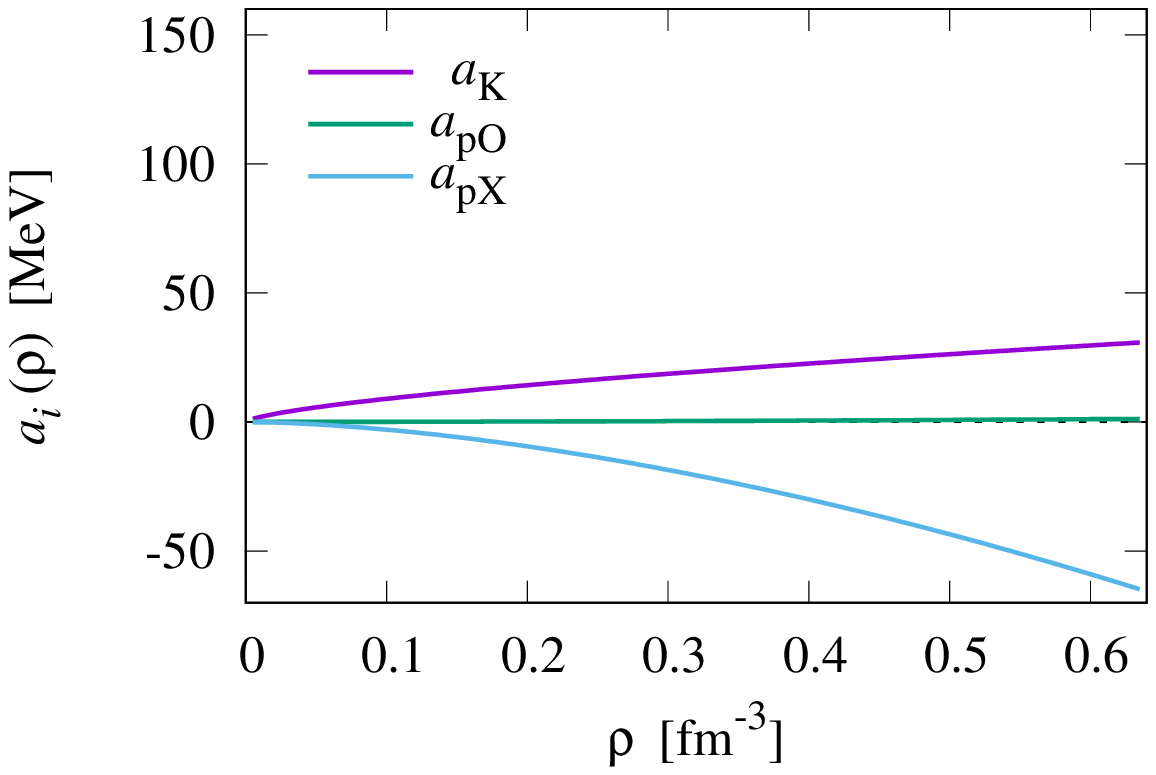}
 \vspace{10pt}
\subcaption{SkM$^\ast$} 
\end{minipage}
 \begin{minipage}[t]{0.5\hsize}
  \centering
  \includegraphics[scale=0.6]{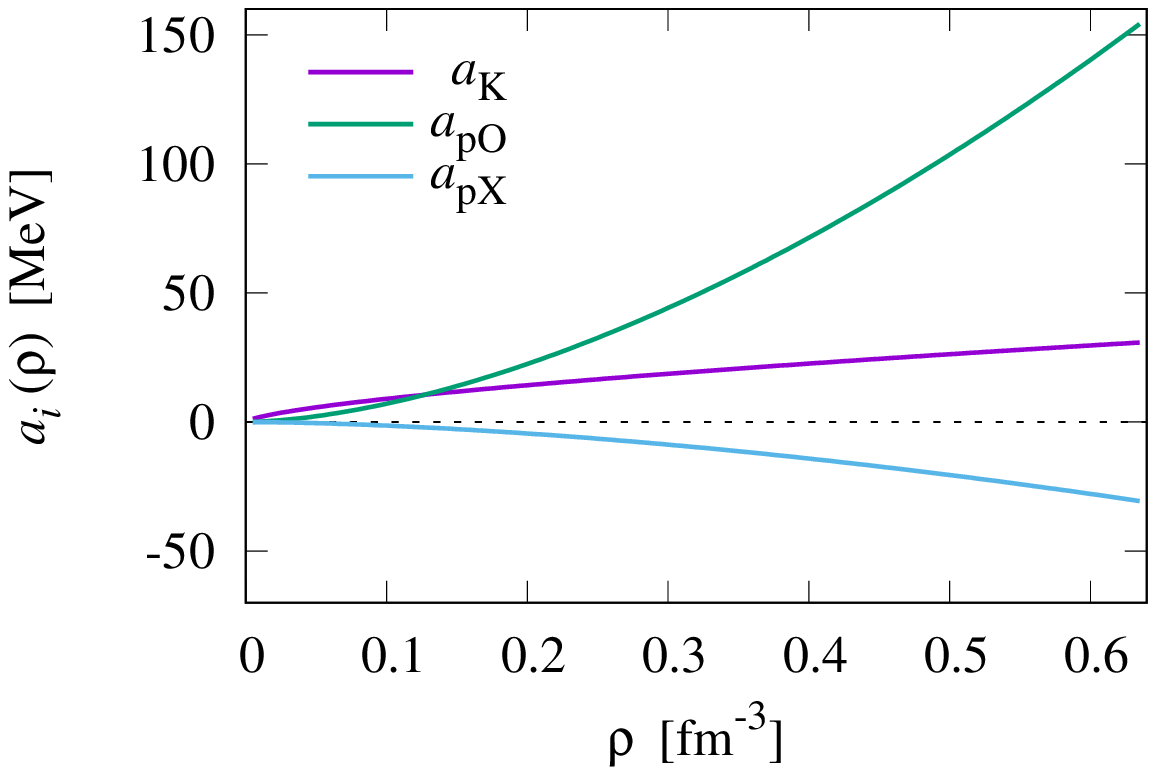}
 \vspace{10pt}
\subcaption{SLy4} 
\end{minipage}
 \caption{Contribution of each term in the symmetry energy
   for (a) SkM$^\ast$ and (b) SLy4.}
\end{figure}
  
Figure~1 accounts for the difference of $\delta_t$
between the Gogny and M3Y-type interactions.
The large $|\Delta_{\mathrm{FX}n}|$ for the Gauss function
at high $k_\mathrm{F}/\mu_n$ makes $|\delta_t|$ large.
Note that $\Delta_t$ in D1S diverges at $\rho=0.55\,\mathrm{fm}^{-3}$
as shown in Fig.~2(a), because $a_t$ reaches $0$.

\section{Summary}

We have investigated accuracy of approximation of the symmetry energy
in the nuclear matter within the Hartree-Fock theory.
As measures of the accuracy of ${\tilde a}_t(\rho)$
(difference of the neutron-matter energy from the symmetric-matter energy)
relative to $a_t(\rho)$
(second derivative of the energy with respect to the asymmetry $\eta_t$),
we have estimated the absolute error $\delta_t$
and the relative error $\Delta_t$.
The errors are decomposed into several terms,
in association with the decomposition of the nuclear-matter energy
in Eq.~\eqref{e}.
With this decomposition,
$\delta_t$ can be expressed in terms of $a_i$ and $\Delta_i$,
where $i$ represents individual components.
We derive analytical expressions for each of $\Delta_i$,
which makes origin of $\delta_t$ and $\Delta_t$ transparent.
On this basis, $\delta_t$ and $\Delta_t$ are estimated
for the six effective interactions by inserting values of the parameters.
We find $-2\,\mathrm{MeV}\lesssim\delta_{t}\lesssim 4\,\mathrm{MeV}$,
which means $|\Delta_t|<0.1$, up to twice of the normal density
for all of the effective interactions.
At higher densities, they depend much on the effective interactions.
The errors is kept small for the two M3Y-type interactions and SLy4,
while the accuracy is worse for the two Gogny interactions and SkM$^\ast$.
We have explained this tendency from $\Delta_i$,
\textit{i.e.}, the relative errors of individual terms in the symmetry energy.

It is recalled that SLy4, D1M, M3Y-P6 and M3Y-P7 are the parameter-sets
which are fitted to the microscopic calculation of the equation-of-state
of the neutron matter.
With an exception of D1M,
this constraint might help keeping the parameter-sets
to have small $\delta_t$ and $\Delta_t$ even at high density. 
For the Skyrme interactions, all of $\Delta_i$ are constant,
and $\delta_t$ can be written by only three terms,
the kinetic term $a_\mathrm{K}$, the momentum-dependent term
between like nucleons $a_\mathrm{pO}$,
and that between unlike nucleons $a_\mathrm{pX}$.
Then, we have found that the parameters of the momentum-dependent terms
$t_\mathrm{p}^{{\rm o}}$ and $t_\mathrm{p}^{{\rm x}}$
make the different behavior of $\delta_t$ between SkM$^\ast$ and SLy4.
For the finite-range interactions,
$\delta_t$ can be expressed by the kinetic term $a_\mathrm{K}$,
the Fock terms between like nucleons $a_{\mathrm{FO}n}$
and that between unlike nucleons $a_{\mathrm{FX}n}$.
For given function types,
the coefficients of the Fock terms $\Delta_{\mathrm{FO}n}$ and $\Delta_{\mathrm{FX}n}$
are single-variable functions of $k_\mathrm{F}/\mu_n$,
which are independent of the parameters.
As shown in Fig.~1,
$\Delta_{\mathrm{FX}n}$ produce significant difference
of $\delta_t$ and $\Delta_t$
between the Gauss and the Yukawa functions at high density.
This suggests that the function type plays a certain role
in the accuracy of approximation.

\section*{Acknowledgment}
We thank K.~Iida for helpful discussions.
This work is financially supported in part by JSPS KAKENHI
Grant Number 24105008 and 16K05342.


%

\appendix
\section{Parameters of interactions}\label{param}

The parameters of the effective interactions used in this paper,
SkM$^\ast$~\cite{SkM*}, SLy4~\cite{SLy4}, D1S~\cite{D1S}, D1M~\cite{D1M},
M3Y-P6 and M3Y-P7~\cite{M3Y-P}, are tabulated in Tables~A1, A2 and A3.

\begin{table}[htbp]
\begin{center}
\label{p-s}
  \caption{Parameters of SLy4 \& SkM$^\ast$.}
\begin{tabular}{|c c|r r|}
  \hline
  \hline
  parameters && SLy4~& SkM$^\ast$~\\
\hline     
$t_0$ & $({\rm MeV\,fm}^3)$ & $-2488.91$ & $-2645.0$ \\
$t_1$ & $({\rm MeV\,fm}^5)$ & $483.13$ & $410.0$ \\
$t_2$ & $({\rm MeV\,fm}^5)$ & $-549.40$ & $-135.0$ \\
$x_0$ & & $0.778$ & $0.090$ \\
$x_1$ & & $0.328$ & $0.000$ \\
$x_2$ & & $-1.000$ & $0.000$ \\
$t_\mathrm{c}^{\rm{o}}$ & $({\rm MeV\,fm}^3)$ & $-2406.95$ & $-382.291$ \\
$t_\mathrm{c}^{\rm{x}}$ & $({\rm MeV\,fm}^3)$ & $-5528.05$ & $-7064.94$ \\
$t_\mathrm{p}^{\rm{o}}$ & $({\rm MeV\,fm}^5)$ & $324.663$ & $5.000$ \\
$t_\mathrm{p}^{\rm{x}}$ & $({\rm MeV\,fm}^5)$ & $575.327$ & $550.00$ \\
$t_\rho^{(\mathrm{SE})}$ & $({\rm MeV\,fm}^{3(1+\alpha)})$ & $-608.49$ & $2599.17$ \\
$t_\rho^{(\mathrm{TE})}$ & $({\rm MeV\,fm}^{3(1+\beta)})$ & $5166.12$ & $2599.17$ \\
$\alpha$ & & $1/6$ & $1/6$ \\
$\beta$ & & $1/6$ & $1/6$ \\\hline
\hline
\end{tabular}
\end{center}
\end{table}

\begin{table}[htbp]
\begin{center}
\label{p-g}
 \caption{Parameter sets of D1S \& D1M.}
\begin{tabular}{|c c|r r|}
  \hline
  \hline
  parameters && D1S~~& D1M~~\\
  \hline
  $\mu_1^{-1}$ & $({\rm fm})$ & $0.7$ & $0.5$ \\  
  $\mu_2^{-1}$ & $({\rm fm})$ & $1.2$ & $1.0$ \\
  $t_1^{(\mathrm{W})}$ & $({\rm MeV})$ & $-1720.3$ & $-12797.57$ \\
  $t_1^{(\mathrm{B})}$ & $({\rm MeV})$ & $1300.0$ & $14048.85$ \\
  $t_1^{(\mathrm{H})}$ & $({\rm MeV})$ & $-1813.53$ & $-15144.43$ \\
  $t_1^{(\mathrm{M})}$ & $({\rm MeV})$ & $1397.6$ & $11963.89$ \\
  $t_2^{(\mathrm{W})}$ & $({\rm MeV})$ & $103.64$ & $490.95$ \\
  $t_2^{(\mathrm{B})}$ & $({\rm MeV})$ & $-163.48$ & $-752.27$ \\ 
  $t_2^{(\mathrm{H})}$ & $({\rm MeV})$ & $162.81$ & $675.12$ \\
  $t_2^{(\mathrm{M})}$ & $({\rm MeV})$ & $-223.93$ & $-693.57$ \\
  $t_\rho^{(\mathrm{SE})}$ & $({\rm MeV\,fm}^{3(1+\alpha)})$ & $0$ & $0$ \\
  $t_\rho^{(\mathrm{TE})}$&$({\rm MeV\,fm}^{3(1+\beta)})$ & $2781.2$ & $3124.44$ \\
  $\alpha$ & & --- & --- \\
  $\beta$  & & $1/3$ & $1/3$ \\
  \hline
  \hline
\end{tabular}
\end{center}
\end{table}

\begin{table} [htbp]
\begin{center} 
\caption{Parameter sets of M3Y-P6 \& M3Y-P7.}
\begin{tabular}{|c c|r r|}
  \hline
  \hline
  Parameters && M3Y-P6 & M3Y-P7 \\
\hline
$\mu_1^{-1}$ & $({\rm fm})$ & $0.25$ & $0.25$ \\
$\mu_2^{-1}$ & $({\rm fm})$ & $0.40$ & $0.40$ \\
$\mu_3^{-1}$ & $({\rm fm})$ & $1.414$ & $1.414$ \\
$t_1^{(\mathrm{SE})}$ & $({\rm MeV})$ & $10766$ & $10655$ \\
$t_1^{(\mathrm{TE})}$ & $({\rm MeV})$ & $8474$  & $9592$  \\
$t_1^{(\mathrm{SO})}$ & $({\rm MeV})$ & $-728$  & $11510$ \\
$t_1^{(\mathrm{TO})}$ & $({\rm MeV})$ & $12453$ & $13507$ \\
$t_2^{(\mathrm{SE})}$ & $({\rm MeV})$ & $-3520$ & $-3556$ \\
$t_2^{(\mathrm{TE})}$ & $({\rm MeV})$ & $-4594$ & $-4594$ \\
$t_2^{(\mathrm{SO})}$ & $({\rm MeV})$ & $1386$  & $1283$  \\
$t_2^{(\mathrm{TO})}$ & $({\rm MeV})$ & $-1588$ & $-1812$ \\
$t_3^{(\mathrm{SE})}$ & $({\rm MeV})$ & $-10.463$ & $-10.463$ \\
$t_3^{(\mathrm{TE})}$ & $({\rm MeV})$ & $-10.463$ & $-10.463$ \\
$t_3^{(\mathrm{SO})}$ & $({\rm MeV})$ & $31.389$  & $31.389$  \\
$t_3^{(\mathrm{TO})}$ & $({\rm MeV})$ & $3.488$   & $3.488$   \\
$t_\rho^{(\mathrm{SE})}$ & $({\rm MeV\,fm}^{3(1+\alpha)})$ & $384$ & $830$ \\
$t_\rho^{(\mathrm{TE})}$ & $({\rm MeV\,fm}^{3(1+\beta)})$ & $1930$ & $1478$ \\
$\alpha$ & & $1$ & $1$ \\
$\beta$ & & $1/3$ & $1/3$ \\
\hline
\hline
\end{tabular}
\end{center}
\end{table}

\section{Formulas for symmetry energy}\label{symeng}

We give some formulas for $a_i$, which is defined in Eq.~\eqref{ai},
in this Appendix.

The second-order derivatives with respect to the asymmetry
of the kinetic-energy term and the density-dependent term are,
\begin{equation}
\begin{aligned}
  {\cal E}_\mathrm{K}^{(2)}
  = \frac{\partial^2{\cal E}_\mathrm{K}}{\partial\eta_t^2}
  &= \frac{\pi^2\rho}{4M}(k_\mathrm{p}^{-1}+k_\mathrm{n}^{-1})\,,\\
  {\cal E}_\mathrm{DD}^{(2)}
  = \frac{\partial^2{\cal E}_\mathrm{DD}}{\partial\eta_t^2}
  &= \frac{1}{8}\bigl(t_\rho^{(\mathrm{SE})}\rho^{\alpha+1}
  -3t_\rho^{(\mathrm{TE})}\rho^{\beta+1}\bigr)\,.
\end{aligned}
\end{equation}
The second-order derivatives of the ${\cal W}$ functions
are represented as follows,
\begin{equation}
\begin{aligned}
  &{\cal W}^{\rm{H}}_n(k_\mathrm{p},k_\mathrm{p})^{(2)}
  + {\cal W}^{\rm{H}}_n(k_\mathrm{n},k_\mathrm{n})^{(2)}
  = 16\pi^6\rho^2\tilde{f}_n(0)\,,\\
  &{\cal W}^{\rm{H}}_n(k_\mathrm{p},k_\mathrm{n})^{(2)}
  =-8\pi^6\rho^2\tilde{f}_n(0)\,,\\
  &{\cal W}^{\rm{F}}_n(k_\mathrm{p},k_\mathrm{p})^{(2)}
  + {\cal W}^{\rm{F}}_n(k_\mathrm{n},k_\mathrm{n})^{(2)}
  =32\pi^6\rho^2\biggl[k_\mathrm{p}^{-4}\int_{0}^{k_\mathrm{p}}dk\,k^3\tilde{f}_n(2k)
  + k_\mathrm{n}^{-4}\int_{0}^{k_\mathrm{n}}dk\,k^{3}\tilde{f}_n(2k)\biggr]\,,\\
  &{\cal W}^{\rm{F}}_n(k_\mathrm{p},k_\mathrm{n})^{(2)}
  =4\pi^6\rho^2\int_{(k_\mathrm{p}-k_\mathrm{n})/2}^{(k_\mathrm{p}+k_\mathrm{n})/2}dk\\
  &\qquad\qquad\qquad \times\,\bigl[
    -\{(k_\mathrm{p}^{-4}+k_\mathrm{n}^{-4})(k_\mathrm{p}^{2}+k_\mathrm{n}^{2})
    +4k_\mathrm{p}^{-1}k_\mathrm{n}^{-1}\} k
    +4(k_\mathrm{p}^{-4}+k_\mathrm{n}^{-4})k^3\bigr]\tilde{f}_n(2k)\,.\\
\end{aligned}
\end{equation}
Each term of Eq.~\eqref{ai} is given by
\begin{equation}
\begin{aligned}
a_\mathrm{K}&=\frac{\pi^2\rho}{4M}k_\mathrm{F}^{-1}\,,\\ 
a_\mathrm{DD}&=\frac{\rho}{16}(t_\rho^{(\mathrm{SE})}\rho^{\alpha +1}
-3t_\rho^{(\mathrm{TE})}\rho^{\beta +1})\,,\\ 
a_{\mathrm{HO}n}&=\frac{1}{2(2\pi)^6}(2t_n^{(\mathrm{W})}+t_n^{(\mathrm{B})}
-2t_n^{(\mathrm{H})}-t_n^{(\mathrm{M})}) 
\left.\bigl\{{\cal W}^{\rm{H}}_n(k_\mathrm{p},k_\mathrm{p})^{(2)}
+ {\cal W}^{\rm{H}}_n(k_\mathrm{n},k_\mathrm{n})^{(2)}\bigr\}\right|_{\eta_t=0}\\
&=\frac{\rho}{8}(2t_n^{(\mathrm{W})}+t_n^{(\mathrm{B})}-2t_n^{(\mathrm{H})}
-t_n^{(\mathrm{M})})\,{\tilde f}_n(0)\,,\\
a_{\mathrm{HX}n}&=\frac{1}{(2\pi)^6}(2t_n^{(\mathrm{W})}+t_n^{(\mathrm{B})})
\left.{\cal W}^{\rm{H}}_n(k_\mathrm{p},k_\mathrm{n})^{(2)}\right|_{\eta_t=0}\\
&=-\frac{\rho}{8}(2t_n^{(\mathrm{W})}+t_n^{(\mathrm{B})})\,{\tilde f}_n(0)\,,\\
a_{\mathrm{FO}n}&=\frac{1}{2(2\pi)^6}(2t_n^{(\mathrm{M})}+t_n^{(\mathrm{H})}
-2t_n^{(\mathrm{B})}-t_n^{(\mathrm{W})}) 
\left.\bigl\{{\cal W}^{\rm{F}}_n(k_\mathrm{p},k_\mathrm{p})^{(2)}
+ {\cal W}^{\rm{F}}_n(k_\mathrm{n},k_\mathrm{n})^{(2)}\bigr\}\right|_{\eta_t=0}\\
&= \frac{\rho}{2}(2t_n^{(\mathrm{M})}+t_n^{(\mathrm{H})}-2t_n^{(\mathrm{B})}
-t_n^{(\mathrm{W})})k_\mathrm{F}^{-4}\int _{0}^{k_\mathrm{F}}dk\,k^{3}\,
  \tilde{f}_n(2k)\,\\ 
a_{\mathrm{FX}n}&=\frac{1}{(2\pi)^6}(2t_n^{(\mathrm{M})}+t_n^{(\mathrm{H})})
\left.{\cal W}^{\rm{F}}_n(k_\mathrm{p},k_\mathrm{n})^{(2)}\right|_{\eta_t=0}\\
&= \frac{\rho}{2}(2t_n^{(\mathrm{M})}+t_n^{(\mathrm{H})})k_\mathrm{F}^{-4}
\int _{0}^{k_\mathrm{F}}dk\,\bigl(-k_\mathrm{F}^2 k+ k^3\bigr)\,\tilde{f}_n(2k)\,.
\end{aligned}
\end{equation}

\section{Explicit expressions of $\Delta_{\mathrm{FO}n}$ and $\Delta_{\mathrm{FX}n}$}
\label{Delta_F}

If the function form is specified,
$\Delta_{\mathrm{FO}n}$ and $\Delta_{\mathrm{FX}n}$ can be expressed
in more explicit manner.
We here denote $k_\mathrm{F}/\mu_n$ by $x$.
Then $\Delta_{\mathrm{FO}n}$ and $\Delta_{\mathrm{FX}n}$ become
functions only of $x$ for a given function form.

For the Gauss function $f(r)=e^{-(\mu r)^2}$,
whose Fourier transform is ${\tilde f}(2k)=(\sqrt{\pi}/\mu)^3\,e^{-(k/\mu)^2})$,
we obtain
\begin{equation}\label{DfG}
\begin{aligned}
  \Delta_{\mathrm{FO}n} &=\frac{3}{x^2}\cdot\frac{1}{1-e^{-x^2}-x^2 e^{-x^2}}
  \Bigl[-1+3(1-2^{-1/3})x^2 + (2-x^2)\,e^{-x^2}\\
    &\qquad -(1-2^{-1/3}x^2)\,e^{-2^{2/3}x^2}+\sqrt{\pi}x^3
    \{{\rm erf}(2^{1/3}x) - {\rm erf}(x)\}\Bigr] -1\,,\\
  \Delta_{\mathrm{FX}n} &=\frac{3}{x^2}\cdot
  \frac{2-3x^2-2e^{-x^2}+x^2 e^{-x^2}+ \sqrt{\pi}x^3\,{\rm erf}(x)}
       {-1+x^2+e^{-x^2}}-1\,,
\end{aligned}
\end{equation} 
where ${\rm erf}(x) = (2/\sqrt{\pi})\int_{0}^{x}e^{-t^2}dt$.

For the Yukawa function $f(r)=e^{-\mu r}/\mu r$,
whose Fourier transform is ${\tilde f}(2k)=4\pi/\mu(\mu^2+4k^2)$,
$\Delta_{\mathrm{FO}n}$ and $\Delta_{\mathrm{FX}n}$ are
\begin{equation}\label{DfY}
\begin{aligned}
\Delta_{\mathrm{FO}n}&=\frac{3}{2x^2}\cdot\frac{1}{\bigl[4x^2-\log(1+4x^2)\bigr]}
\Bigl[2(1-2^{-1/3})x^2 + 12(2^{1/3}-1)x^4 \\
  &\qquad\qquad + \frac{1}{4}\log\frac{1+4\cdot2^{2/3}x^2}{(1+4x^2)^2}
  +3\cdot 2^{2/3}x^2\log(1+4\cdot 2^{2/3}x^2) -6x^2\log(1+4x^{2})\\
  &\qquad\qquad -16x^3\{{\rm arctan}(2^{4/3}x) - {\rm arctan}(2x) \}
  \Bigr] - 1\,,\\
\Delta_{\mathrm{FX}n}&=\frac{3}{4x^2}\cdot\frac{4x^2(1-6x^2)+32x^3{\rm arctan}(2x)
  -(1+12x^2)\log(1+4x^2)}{4x^2-(1+4x^2)\log(1+4x^2)}-1\,.
\end{aligned}
\end{equation}

\end{document}